\documentclass[sigconf,nonacm]{acmart}

\usepackage{booktabs}
\usepackage{tabularx}
\usepackage{tabu}                      
\usepackage{booktabs}                  
\usepackage{lipsum}                    
\usepackage{mwe}                       
\usepackage[most]{tcolorbox}
\usepackage{placeins}
\usepackage{float}

\definecolor{claimagency}{HTML}{E36863}
\definecolor{claimadapt}{HTML}{006BBB}
\definecolor{claimtransp}{HTML}{18A87C}

\newcommand{\claimtagagency}[1]{\textcolor{claimagency}{\bfseries #1}\ }
\newcommand{\claimtagadapt}[1]{\textcolor{claimadapt}{\bfseries #1}\ }
\newcommand{\claimtagtransp}[1]{\textcolor{claimtransp}{\bfseries #1}\ }

\newtcolorbox{claimboxagency}[1][]{%
  enhanced,
  breakable,
  colback=white,
  colframe=white,
  boxrule=0pt,
  sharp corners,
  borderline west={2pt}{0pt}{claimagency},
  left=2pt,right=4pt,top=-2pt,bottom=-2pt,
  before skip=5pt,
  after skip=5pt,
  parbox=false,
  #1
}
\newtcolorbox{claimboxadapt}[1][]{%
  enhanced,
  breakable,
  colback=white,
  colframe=white,
  boxrule=0pt,
  sharp corners,
  borderline west={2pt}{0pt}{claimadapt},
  left=2pt,right=4pt,top=-2pt,bottom=-2pt,
  before skip=5pt,
  after skip=5pt,
  parbox=false,
  #1
}
\newtcolorbox{claimboxtransp}[1][]{%
  enhanced,
  breakable,
  colback=white,
  colframe=white,
  boxrule=0pt,
  sharp corners,
  borderline west={2pt}{0pt}{claimtransp},
  left=2pt,right=4pt,top=-2pt,bottom=-2pt,
  before skip=5pt,
  after skip=5pt,
  parbox=false,
  #1
}

\AtBeginDocument{%
  }

\setcopyright{none}
\copyrightyear{2026}
\acmYear{2026}
\acmDOI{}
\acmISBN{}

\acmConference[CSCW '26 Companion]{Companion of the Computer-Supported
  Cooperative Work and Social Computing}{October 10, 2026}{Salt Lake City, UT, USA}
\acmBooktitle{Workshop on Broader Impacts of GenAI in Communication:
  Building Agendas for Research, Design, and Policymaking
  (CSCW '26 Companion), October 10, 2026, Salt Lake City, UT, USA}

\begin{document}

\title{Sensemaking as Artifact: Accumulated Influence in AI-Mediated Information Environments}

\author{Manling Yang}
\affiliation{%
  \institution{Tufts University}
  \city{Medford}
  \state{MA}
  \country{USA}
}

\author{Remco Chang}
\affiliation{%
  \institution{Tufts University}
  \city{Medford}
  \state{MA}
  \country{USA}
}

\renewcommand{\shortauthors}{Yang and Chang}

\begin{abstract}
Generative AI is changing what can happen after a source artifact reaches its audience. A viewer’s interpretation can now be externalized into a derivative artifact, allowing private sensemaking to become part of subsequent communication. Once such a derivative artifact circulates, it can enter subsequent viewers’ information environments and shape the conditions under which their later sensemaking occurs.
In this paper, we examine how this shift changes visual information communication. We first consider the viewer’s immediate interaction with a source artifact and generative AI. We then examine what becomes consequential when the viewer’s sensemaking takes communicative form, including communicative commitment, the legibility of transformations and source relationships, and the literacy required to interpret already-mediated information. Finally, we broaden the unit of analysis to consider how repeated and distributed AI mediation may accumulate over time, shaping what subsequent viewers notice, consider plausible, trust, and carry into subsequent sensemaking. We argue that understanding these longer-term forms of influence is a research direction for AI-mediated visual communication.
\end{abstract}

\ccsdesc[500]{Human-centered computing~Visualization}
\ccsdesc[500]{Human-centered computing~Collaborative and social computing}
\ccsdesc[300]{Computing methodologies~Artificial intelligence}

\keywords{AI-mediated communication, visual information communication}

\maketitle

\section{Introduction}
AI-mediated communication (AIMC) describes communication in which an AI system modifies, augments, or generates messages as part of an exchange~\cite{hancock2020aimc}. Recent work suggests that this mediation is becoming increasingly ordinary as general-purpose generative AI systems are incorporated into everyday information seeking and sensemaking~\cite{peng2026lay,zhou2026understanding}. This development broadens the role of AI in communication beyond assisting a designer in composing a message. AI can also enter at reception, after a source artifact has reached its audience, and participate in how that artifact is examined and understood. For visual communication, this means that a source representation need not remain the only representational form through which its information is encountered. A viewer may bring the source artifact into an AI-mediated exchange, where its content can be explained, contextualized, questioned, and subsequently represented in new forms. This extension of mediation into reception motivates us to consider what happens to visual communication after a source artifact has been handed off to its viewer.

This shift creates new opportunities for visual communication. Generative AI can make a source artifact more accessible without requiring its designer or source organization to anticipate the needs of every viewer. For example, a viewer of a medical research paper may use AI to clarify unfamiliar terminology or request an explanation at an appropriate level of detail~\cite{august2023paperplain}. Similar capabilities can support personalization and adaptation across viewers~\cite{kirk2024personalization,yanez2025state,shin2025drillboards}. A viewer can engage with a source artifact around their own questions, goals, and context by asking for alternative explanations, reorganizations, or reformulations of its content.

However, the same responsiveness that enables personalized sensemaking also means that these interpretations need not be neutral representations of the source information. An interpretation can reflect not only the source artifact but also the viewer's goals, context, and existing perspective. When AI systems have access to prior interaction context, this influence can extend beyond the immediate encounter: their responses can reflect perspectives inferred from earlier exchanges, e.g., through sycophantic behavior~\cite{jain2026sycophancy}. At the same time, generative AI is becoming part of how people produce content for others: studies of content creators document its use across platforms and throughout creative workflows, while social media platforms increasingly confront the circulation of AI-generated and AI-co-created content~\cite{Kim2026content,anderson2025making,gao2026governance}. Together, these developments create conditions in which sensemaking situated in a viewer's goals, context, and prior interactions with AI can be externalized into a derivative artifact and subsequently encountered by other viewers.

A derivative artifact may preserve elements of the source artifact (e.g., data, visual conventions, source attribution, etc.) while reframing its emphasis or claim. Such retained source and authority cues can shape how subsequent viewers assess the credibility of the derivative artifact~\cite{sundar2008main,lin2016social}. A derivative artifact may therefore appear to carry the authority of the original designer or source organization even when its communicative claim has changed. As derivative artifacts circulate and undergo further AI-mediated transformation, their transformation history may also become harder to recover. Subsequent viewers may no longer know where a claim originated, what aspects of the artifact have been changed, or who was responsible for those changes. Such provenance information can be consequential for how audiences assess digital media~\cite{feng2023examining,francis2023transparency}. More broadly, once derivative artifacts circulate, they become part of the information available to subsequent viewers. What began as one viewer's situated sensemaking can therefore shape the conditions under which subsequent viewers interpret information. This raises questions not only about how source and derivative artifacts are received and transformed, but also about how AI-mediated sensemaking becomes part of the information environments that shape subsequent interpretation.

\begin{figure*}[t]
    \centering
    \includegraphics[width=\textwidth]{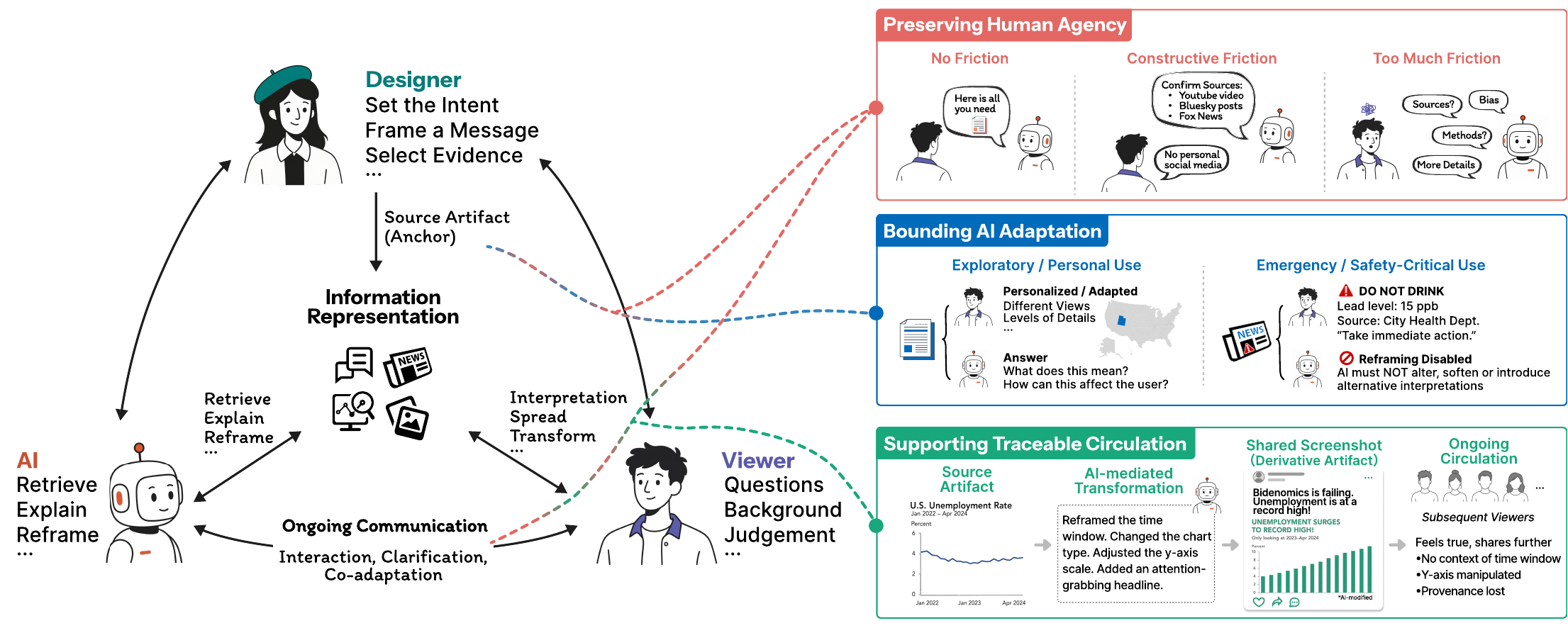}
    \caption{AI-mediated visual communication during and after reception. A viewer can use generative AI to retrieve, explain, reframe, and transform an existing representation. The right side highlights three immediate concerns within this viewer-side encounter: human agency, the scope of AI adaptation, and the visibility of transformation context as interpretations are shared and circulated.}
    \label{fig:framework}
    \vspace{-0.2em}
\end{figure*}

\section{Reception Becomes Actionable}
\label{sec:handoff}
Viewers can engage generative AI with visual information for different purposes and at different levels of intervention. They may seek additional explanation or context, incorporate external information, or reorganize and recreate a source artifact around a particular question. Prior work has explored personalization across audiences~\cite{kirk2024personalization} and natural-language modification of visualizations~\cite{vaithilingam2024dynavis}. In this section, we focus on how these viewer-side interactions reshape visual communication when interpretation is externalized into a derivative artifact and becomes available to subsequent viewers.
Figure~\ref{fig:framework} summarizes the viewer-side encounter we consider in this section. Rather than treating reception as the endpoint of communication, we examine how AI-mediated interaction can alter the viewer’s role, the permissible scope of adaptation, and the visibility of prior transformations once an interpretation takes communicative form. The following subsections unpack these three concerns.

\begin{claimboxagency}
\claimtagagency{Human agency during AI-mediated interpretation.}\\
AI can support a viewer's interpretation while also changing when and how the viewer exercises judgment.
\end{claimboxagency}
\noindent
AI can supply a fluent interpretation before a viewer has examined the basis of the AI-generated account, creating conditions for overreliance on AI recommendations~\cite{buccinca2021trust}. Sycophantic responses introduce a related concern by reinforcing a viewer's prior judgment while increasing trust in the AI system~\cite{cheng2026sycophantic}. The question for interface design is therefore not simply whether viewers retain control, but when the interaction requires them to exercise judgment. Prior work suggests that exposing intermediate steps can improve transparency and controllability~\cite{wu2022aichains}, while cognitive forcing functions can create deliberate moments of choice before accepting AI recommendations~\cite{buccinca2021trust}. Such interventions need not appear at every step. Repeated verification can make interaction burdensome, whereas introducing friction around judgments that substantially alter the message or inform subsequent action may create more consequential opportunities for reflection.

\begin{claimboxadapt}
\claimtagadapt{Variation in the scope of AI adaptation.}\\
The consequences of adaptation depend on what part of a source artifact changes and on the context in which that artifact operates.
\end{claimboxadapt}
\noindent
AI-mediated adaptation can make information more accessible by adjusting its explanation, level of detail, or presentation to a viewer's needs~\cite{august2023paperplain,yanez2025state}. Yet not all parts of a source artifact are equally open to adaptation. The lead notice in Figure~\ref{fig:framework} illustrates this distinction. An AI system might explain what a lead measurement means for a particular household while preserving ``DO NOT DRINK'' as an instruction issued by the source organization. Changing the explanation adapts how the information is understood, while changing the instruction may alter what the source organization is asking the viewer to do. The appropriate scope of adaptation therefore depends not only on what AI is capable of changing, but also on the communicative function of the content and the stakes of the setting.

\begin{claimboxtransp}
\claimtagtransp{Transformation context during sharing.}\\
When an AI-mediated interpretation takes a shareable form, information about how it was produced may no longer remain visible alongside it.
\end{claimboxtransp}
\noindent
A viewer may leave an AI-mediated exchange with a screenshot, rewritten post, or derivative artifact. The example in Figure~\ref{fig:framework} changes the visible time window, chart type, y-axis scale, and headline of an unemployment chart. Such changes can substantially alter the interpretation invited by a visualization even when the underlying data remain unchanged~\cite{pandey2015deceptive,correll2020truncating,lo2022misinformed}. Once the derivative artifact is shared, however, subsequent viewers may encounter it without access to the questions, prompts, or transformation choices through which it was produced. Prior work on digital media provenance shows that information about authorship and edit history can affect how audiences assess altered media~\cite{feng2023examining,francis2023transparency}. A generic ``AI-generated'' or ``AI-modified'' label may identify the presence of AI mediation, but it does not make this transformation context legible: what was changed, why it was changed, or how the derivative artifact relates to its source artifact.

These concerns become more consequential once a viewer’s interpretation takes communicative form and becomes available to subsequent viewers, which we consider next.
\section{Sensemaking Becomes Communicative}
\label{sec:rethink}
Once an interpretation takes communicative form, the design problem extends beyond the viewer's immediate encounter. We consider three questions: how systems can make communicative commitment more explicit (Section~\ref{sec:communicative-role}), how transformations and source relationships can remain legible (Section~\ref{sec:transformation-legibility}), and what literacy subsequent viewers need when another viewer's sensemaking enters their information environment (Section~\ref{sec:literacy-mediated}).

\subsection{Communicative Commitment}
\label{sec:communicative-role}

Once an AI-mediated interpretation is shared, choices made during sensemaking acquire a different status. Exploring an interpretation does not necessarily imply endorsing it, but carrying that interpretation into a derivative artifact for subsequent viewers turns selected choices into part of a communicative commitment. The design problem is therefore to make this transition legible without requiring the full private interaction to be exposed. For the viewer, systems could make consequential choices more explicit at the point of externalization, for example by distinguishing tentative exploration from content that will be carried into a derivative artifact. Prior work on seamful design suggests ways of making consequential aspects of AI-mediated interaction visible~\cite{ehsan2024seamful}, while cognitive forcing functions demonstrate how interfaces can introduce deliberate moments of reflection around AI-supported judgments~\cite{buccinca2021trust}. These ideas motivate selective intervention at the point where private sensemaking becomes communication.

For subsequent viewers, the corresponding problem is to understand the status of the derivative artifact. AI mediation can complicate how subsequent viewers attribute agency and responsibility for communicated content~\cite{hohenstein2020ai}. A derivative artifact may reproduce the source artifact, contextualize it, or advance a revised account, yet these relationships may not be apparent from the derivative artifact alone. Making communicative commitment legible therefore requires supporting both sides of the exchange: helping the viewer recognize what they are carrying forward and helping subsequent viewers understand the position from which the derivative artifact is presented.

\subsection{Legibility Across Transformation}
\label{sec:transformation-legibility}

Once a derivative artifact is shared, making its relationship to the source artifact legible requires more than indicating that a transformation occurred. A generic label such as ``AI-modified'' identifies the presence of mediation but says little about its communicative significance. At the other extreme, simply exposing more information does not necessarily make an AI-mediated process more intelligible, as different levels of transparency can produce different effects on users' understanding and trust~\cite{kizilcec2016much}. Preserving a complete history of prompts, responses, and intermediate representations may therefore provide extensive provenance without clarifying which changes matter for understanding the derivative artifact. The design problem is to identify and represent transformations that meaningfully change how the source artifact is presented to subsequent viewers.

Such transformations can differ not only in what they change, but also in how they alter the derivative artifact's relationship to its source artifact. Translating a warning, adding contextual information, inferring a new claim, and revising a recommendation each preserve different aspects of the original communication. This distinction becomes particularly important when recognizable source cues remain. Institutional identity and other authority cues can shape perceived credibility~\cite{lin2016social,sundar1998effect}, while provenance research shows that the presentation of authorship and edit history can affect how audiences assess altered media~\cite{feng2023examining}. A derivative artifact may therefore remain correctly attributed to its source while presenting claims that the original designer or source organization did not make or endorse.

Designing for legibility requires connecting information about transformation to its communicative consequence. Rather than treating attribution as something to preserve or remove from the derivative artifact as a whole, systems could distinguish content that is retained, translated, contextualized, inferred, or revised, and communicate how these changes affect its relationship to the source artifact. In the public-health example in Figure~\ref{fig:framework}, the health department can remain the source organization of the underlying warning after translation or explanation, while a subsequently revised recommendation should not automatically inherit the same institutional endorsement. More generally, provenance may need to communicate not only where a derivative artifact came from and how it changed, but which parts of the resulting communication can still be understood as attributable to the original source.

\subsection{Literacy for AI-Mediated Information}
\label{sec:literacy-mediated}

\begin{figure}[H]
    \centering
    \vspace{-0.5em}
    \includegraphics[width=0.95\columnwidth]{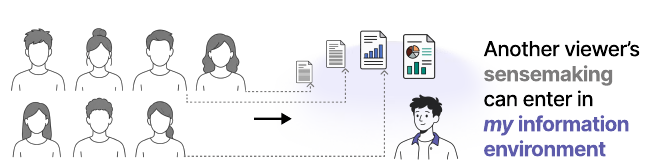}
    \caption{One viewer's sensemaking can become part of a subsequent viewer's information environment, shaping the information available for subsequent interpretation.}
    \label{fig:information-environment}
    \vspace{-0.5em}
\end{figure}

As illustrated in Figure~\ref{fig:information-environment}, one viewer's sensemaking can become part of a subsequent viewer's information environment. Evaluating a derivative artifact requires more than understanding its visible content. A subsequent viewer may need to reason about how the source artifact has been selected, reframed, supplemented, or transformed, and how those interventions affect what can be inferred from the derivative artifact. Existing forms of data and visualization literacy concern people's ability to read, interpret, and construct data representations~\cite{borner2019data}, while AI literacy includes competencies for understanding, interacting with, and critically evaluating AI systems~\cite{long2020ai}. In AI-mediated visual communication, these forms of literacy meet in a relational problem: understanding how the source artifact, the viewer's interpretation, and AI mediation jointly shape the derivative artifact.

Making mediation legible is therefore necessary but not sufficient. A transformation can be explicitly disclosed without making its significance obvious to a subsequent viewer. Prior work similarly shows that the effects of transparency depend on what information is presented and how~\cite{kizilcec2016much}. For example, knowing that a y-axis scale was truncated does not by itself explain how that choice changes the comparison invited by a chart~\cite{correll2020truncating}. Similarly, knowing that a claim was AI-generated does not determine how it should be weighed against source evidence. Design should therefore consider not only what information about mediation is exposed, but how subsequent viewers can use that information in their own judgment. This opens questions about what competencies are needed to distinguish source evidence from the viewer's prior interpretation, assess the consequences of particular transformations, and recognize when an apparently self-contained derivative artifact already reflects another person's situated sensemaking.

This perspective also changes the scale of the problem. A subsequent viewer rarely encounters only one derivative artifact. One viewer's sensemaking can become part of the information environment in which a subsequent viewer's interpretation occurs, alongside artifacts shaped by other people, systems, and sources. This broader view is consistent with distributed cognition accounts that treat cognitive activity as extending across people, artifacts, and environments~\cite{hollan2000distributed}. The challenge therefore extends beyond literacy for any single artifact toward understanding how people reason within information environments increasingly populated by the externalized sensemaking of others. This provides the basis for considering how such influence may accumulate over time, which we turn to in Section~\ref{sec:conclusion}.
\section{Accumulated Influence}
\label{sec:conclusion}
The preceding sections considered AI-mediated visual communication first within a viewer's immediate encounter and then as the viewer's sensemaking becomes communicative and enters the information available to subsequent viewers. Here, we broaden the unit of analysis to consider how these processes may accumulate over time.

Generative AI may not only affect how a particular source or derivative artifact is interpreted, but also repeatedly participate in the conditions under which people encounter and make sense of information.
Emerging work already suggests that AI-mediated interaction can have consequences for judgments and intentions beyond the content of an immediate response. For example, sycophantic systems can reinforce viewers' prior judgments and increase confidence in those judgments~\cite{cheng2026sycophantic,jain2026sycophancy}.
This raises a broader question for AI-mediated communication: how might repeated and often subtle acts of mediation accumulate to shape what people notice, consider plausible, trust, and carry into subsequent sensemaking?

\begin{figure}[H]
    \centering
    \vspace{-0.5em}
    \includegraphics[width=0.95\columnwidth]{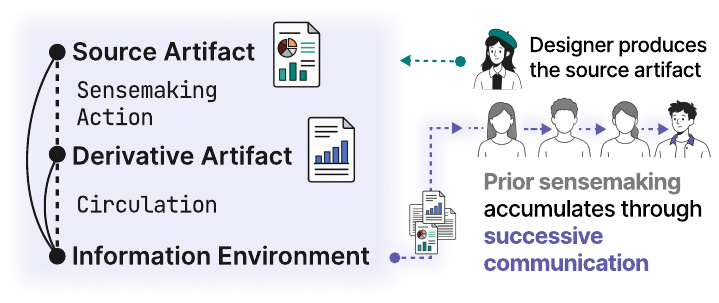}
    \caption{One pathway through which prior sensemaking can accumulate across successive AI-mediated communication.}
    \label{fig:discussion}
    \vspace{-0.5em}
\end{figure}

Importantly, such influence need not come only from a viewer's direct interactions with AI. Information environments can increasingly contain the externalized sensemaking of other viewers, including derivative artifacts that have already been selected, explained, reframed, or supplemented through AI mediation. As illustrated in Figure~\ref{fig:discussion}, one viewer's sensemaking can be externalized into a derivative artifact, enter a subsequent viewer's information environment, and become a basis for further interpretation and action. Prior work on person-to-person transmission shows that repeated retelling can progressively reshape the framing environment itself~\cite{coronel2023competitive}. Through successive AI-mediated communication, particular framings, assumptions, or claims may similarly persist and shape what subsequent viewers encounter.

These processes motivate a research agenda on the long-term cognitive consequences of AI-mediated information environments. Research could examine how repeated AI mediation interacts with existing beliefs and information-seeking behavior, which framings become reinforced across encounters, and whether individually reasonable acts of explanation, personalization, and transformation can collectively narrow what people attend to or consider plausible. Emerging evidence that generative AI can improve individual creative output while reducing collective diversity provides one precedent~\cite{doshi2024generative}. Such effects may be hard to recognize when they emerge gradually across many systems, derivative artifacts, and other viewers' AI-mediated sensemaking rather than from a single source artifact. Communication research has long considered how media exposure and subsequent selection can recursively reinforce one another~\cite{slater2007reinforcing}. Understanding accumulated influence therefore requires moving beyond isolated interactions to study how AI-mediated information environments develop over time and shape subsequent human sensemaking.
\section{Conclusion}
Generative AI extends visual communication beyond the handoff of a source artifact by making reception itself actionable. Viewers can interpret and transform source artifacts around their own goals and contexts, and these interpretations can be externalized into derivative artifacts that enter subsequent viewers’ information environments. These changes mean that the influence of AI-supported sensemaking can extend beyond any single viewer, artifact, or interaction. Understanding AI-mediated visual communication requires attending not only to how AI supports individual interpretation, but also to how those interpretations become communicative and how their influence may accumulate over time.

\bibliographystyle{ACM-Reference-Format}
\bibliography{references}

\end{document}